\documentclass[nofootinbib,
 reprint,
 amsmath,amssymb,
 aps,prl
]{revtex4-2}

\usepackage{graphicx}
\usepackage[english]{babel}
\usepackage{dcolumn}
\usepackage{bm}
\newcommand{\ud}{\mathrm{d}}
\newcommand{\ue}{\mathrm{e}}
\newcommand{\ui}{\mathrm{i}}

\begin{document}


\title{Signature of Einstein-Cartan theory}

\author{B. Arderucio Costa}
\email{bruno.arderucio@correo.nucleares.unam.mx}
\author{Y. Bonder}
 \email{bonder@nucleares.unam.mx}
\affiliation{
 Instituto de Ciencias Nucleares, Universidad Nacional Autonoma de Mexico\\
 Apartado Postal 70-543, Cd.~Mx., 04510, Mexico
}

\date{\today}

\begin{abstract}
We study the effects of torsion as predicted by the Einstein-Cartan theory in the test-particle non-relativist approximation. We derive the corresponding 2-spinor Hamiltonian. Then, we solve an idealized reflection and transmission problem for a beam travelling across a spin-polarized target. We identify deviations in the spin polarizations on the reflected and transmitted beams that can distinguish Einstein-Cartan from general relativity. These deviations would constitute compelling evidence for a non-trivial spacetime torsion if measured.
\end{abstract}

\maketitle

\section{Introduction}
General Relativity (GR) has passed all empirical tests at Solar system scales \cite{will_2018} and is the paradigmatic theory of gravity. Still, modified gravity theories are put forward to improve our description of cosmological phenomena~\cite{CLIFTON20121}. One of the earliest modified gravity theories ---dating back 100 years--- goes by the name of Einstein--Cartan theory (EC)~\cite{cartan1923,cartan1924} and can be derived by requiring invariance under the local Poincar\'e group~\cite{Kibble, Sciama} (for a review see Ref.~\cite{hehl76}).

In EC, gravity is described by a metric (or a tetrad) and a torsion-full connection. The EC action is symbolically identical to the Einstein--Hilbert action of GR, but, in the former, the Ricci curvature depends on the torsion-full connection. As a result, torsion is linked algebraically with the so-called spin density [see Eq.~\eqref{eceom} below]. Consequently, torsion does not propagate and is only non-zero inside matter with spin density. Moreover, the theory reduces to GR whenever torsion vanishes. Thus, vacuum tests, as occurs effectively for most GR experiments, cannot empirically distinguish between GR and EC. Still, knowing the fundamental gravitational degrees of freedom is of the utmost relevance, particularly when attempting to quantize gravity. Of course, one can construct other theories with torsion~\cite{AlternativeModels}. However, the fact that EC reproduces the GR phenomenology in the torsion-free limit suggests that any empirically adequate theory should contain EC as the dominant contribution.

Our action variables are the tetrad $e_a^\mu$, namely a set of orthonormal (dual) vector fields, and an independent spin connection $\omega_{a\mu\nu}$\footnote{We work in $4$ spacetime dimensions and in units where $c=1=\hbar$. Indices for tangent space are denoted by Greek letters, and spacetime indices are indicated by Latin letters from the beginning of the alphabet. Repeated indices imply contraction, and Greek indices are lowered (raised) using $\eta_{\mu\nu}={\rm diag}(-1, 1,1,1)$ ($\eta^{\mu\nu}$, the inverse matrix of $\eta_{\mu\nu}$). Latin indices $i,j,k$ are used for spatial Minkowskian coordinates and, when no confusion arises, are also used for the corresponding tangent space elements.}. The former contains the metric information, and the latter also encodes the torsional degrees of freedom.

The variation of the EC action with respect to $\omega_{a\mu\nu}$ yields
\begin{equation}
{T^\rho}_{bc}(e_\rho^ae^b_\mu e^c_\nu + e^a_\mu e^b_\nu e_\rho^c  + e^a_\nu e^b_\rho e^c_\mu)= 8\pi G {\Sigma^a}_{\mu\nu},
    \label{eceom}
\end{equation}
where ${T^\mu}_{ab}$ is the torsion tensor, ${\Sigma^a}_{\mu\nu} \equiv 2 \delta \mathcal{L}_{\rm M}/\delta {\omega_a}^{\mu\nu}$ is the spin density, $\mathcal{L}_{\rm M}$ is the matter Lagrange function, and $G$ stands for Newton's gravitational constant. On the other hand, the action variation with respect to $e_a^\mu$ generates an Einstein-like equation for a torsion-full Einstein tensor. Relevantly, one can use Eq.~\eqref{eceom} to replace torsion in this equation in terms of ${\Sigma^a}_{\mu\nu}$; this equation differs from the conventional Einstein equation by terms of order $G^2$~\cite{hehl76}.

In the Standard Model of particle physics, the only action terms that couple to the spin connection are those associated with spinors\footnote{Gamma matrices satisfy $\{\gamma^\mu,\gamma^\nu\}=-2\eta^{\mu\nu}$, and $\gamma_5\equiv \ui \gamma^0\gamma^1\gamma^2\gamma^3$. The Dirac adjoint is $\bar\Psi=\Psi^\dagger\gamma^0$. We adopt the Dirac representation for the gamma matrices (i.e., as in Ref.~\cite{bjorken64}), and $\sigma^i$ denotes the usual Pauli matrices.}. Thus, we take $\mathcal{L}_{\rm M}$ as the Lagrangian for a spin-$\frac{1}{2}$ Dirac field $\Psi$, which contains~\cite[ch. 7.10.3]{Nakahara}
\begin{equation}
   \nabla_a\Psi=\left(\partial_a+\frac{1}{4}\omega_{a\mu\nu}\gamma^{\mu\nu}\right)\Psi,\qquad  \gamma^{\mu\nu}\equiv \frac{[\gamma^\mu,\gamma^\nu]}{2}.\label{covpsi}
\end{equation}
In this case Eq.~\eqref{eceom} becomes
\begin{equation}
    {T^\rho}_{\mu\nu}=4\pi G{\epsilon_{\mu\nu}}^{\rho\sigma}{J_5}_\sigma,
    \label{ecdeq}
\end{equation}
where ${T^\rho}_{\mu\nu}$ are the torsion components in the $e_a^\mu$ basis, $\epsilon_{\mu\nu\rho\sigma}$ is the completely anti-symmetric tensor (such that $\epsilon_{0123}=1)$
and $J_5^\mu\equiv \bar\Psi\gamma_5\gamma^\mu\Psi$ is the source's axial current, which is spacelike\footnote{This was pointed out by an anonymous referee.}~\cite{Crawford}. Notice that only the completely anti-symmetric part of the torsion tensor has physical effects. This is a well-known feature of the EC theory~\cite{hehl76}. Also, one recovers GR whenever ${J_5}_\mu=0$.

In this Letter, we compare a reflection and transmission problem for a test spinor in GR and EC when traversing a target with $J_5^\mu\neq 0$. Concretely, we compute the spin polarization angle deflection for a non-relativistic beam of polarized neutrons. There are proposals to search for spin-dependent gravitational effects~\cite{ReviewEffectsSpin}. However, most of those looking for torsion allow for its propagation~\cite{Modified79, PhysRevD.25.573, ModTheos94,PhysRevLett.103.261801, ModTheos14, ModTheos21, ModTheos22}, or ignore the torsion source~\cite{Background97, Background98, Background08, PolarizedTargetGeneralParam, Background15, Background21}, sometimes considering torsion parts that vanish in EC. In addition, there are tests of EC in situations with little experimental control, such as particle collisions~\cite{Acel07, Acel10, Acel14}. Moreover, there is a programme to look for torsion sourced by classical angular momentum~\cite{PhysRevD.76.104029}, which has been disputed~\cite{PhysRevD.75.124016, HEHL20131775}.

The present analysis has three main advantages when compared with existing proposals: first, it focuses on EC, which, as we argue above, is expected to provide the dominant contribution in any theory that reduces to GR. Second, torsion is the only modification to conventional physics, and third, even though the torsion interaction occurs when the beam is inside the target, the detectors may be placed outside, which offers an enormous practical advantage.

\section{Schr\"odinger-like equation}

In this section, we study the evolution of a test Dirac spinor $\psi$ in EC theory, minimally coupled to conventional gravity and torsion; this theory is sometimes called Einstein--Cartan--Dirac theory. We then obtain a non-relativistic approximation for a beam of such particles in the presence of a target. The equation of motion for $\psi$ reads~\cite{Yuri}
\begin{equation}
\ui\gamma^\mu e^a_\mu\left(\partial_a+\Gamma_a\right)\psi-\frac{3 \pi G}{8}J_5^\mu\gamma_5\gamma_\mu\psi=m\psi,
    \label{diraco}
\end{equation}
where $\Gamma_a\equiv \mathring{\omega}_{a\mu\nu}\gamma^{\mu\nu}/4$ and $\mathring{\omega}_{a\mu\nu}$ is the GR connection (which is torsion independent) and $J_5^\mu$ is the target's axial current. To get Eq.~\eqref{diraco}, we used Eq.~\eqref{ecdeq} and the identity $\{\gamma_\rho,\gamma_{\mu\nu}\}=-2\ui\epsilon_{\rho\mu\nu\sigma}\gamma_5\gamma^\sigma$. We also employed the test particle approximation, neglecting the effects of $\psi$ on curvature and torsion. Notice that GR only differs from EC in the last term on the left-hand side of Eq.~\eqref{diraco}.

We work in the weak-field regime for gravity, keeping only linear terms in $G$. The metric is linearized around the Minkowski metric $\eta_{ab}$ as $g_{ab}=\eta_{ab}+G h_{ab}$, where $h_{ab}$ is found by solving the equations of motion. The GR spin connection is $\mathcal{O}(G)$:
\begin{equation}
    \mathring{\omega}_{a\mu\nu}\approx- G e^\rho_a(\partial_\mu h_{\nu\rho}-\partial_\nu h_{\mu\rho}).
    \label{linconn}
\end{equation}

\begin{figure}[ht]
    \centering
    \includegraphics[width=.75\columnwidth]{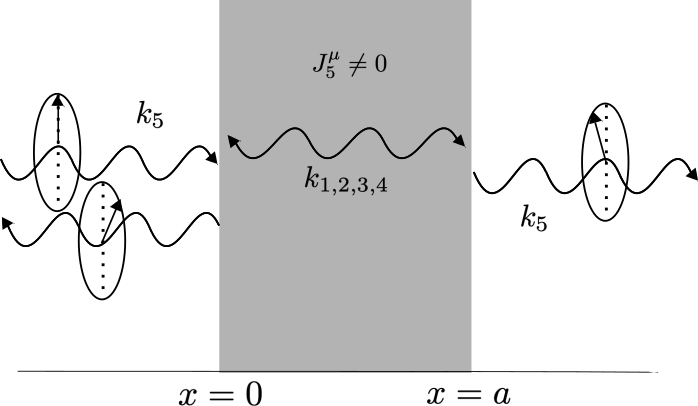}
    \caption{Set-up and auxiliary constructions. The boundary conditions consist of a right-moving incident beam, whose polarization (schematically represented in the wheels) is aligned with the $z$ axis, and a purely right-moving transmitted beam from a slab occupying the $0\leq x\leq a$.
    }
    \label{fig}
\end{figure}

Also, we consider a static target that, for all practical purposes, can be regarded as infinite in the two orthogonal directions, $y$ and $z$, and uniform between the $x=0$ and $x=a>0$ planes, as depicted in Fig.~\ref{fig}. We choose Minkowski coordinates in the target's rest frame compatible with its symmetries. Assuming that the metric perturbations go to zero at long distances from the (static) source, they can be chosen to be diagonal and time-independent\footnote{The linearized Einstein equations for $\bar h_{ab}\equiv h_{ab}-\frac{1}{2}\eta_{ab} h^c_c$ are $\partial_c\partial^c\bar h_{ab}=-16\pi T_{ab}$, where $T_{ab}$ is the target's energy-momentum tensor, and its solutions are the convolution between $T_{ab}$ and the Green function for the flat-space wave operator. It follows from staticity that $h_{0i}=0$ and, assuming the target is shear-free (i.e., the off-diagonal elements of $T_{ij}$ are zero), $h_{ij}=0$ for $i\neq j$.}. In fact, according to Eq.~\eqref{linconn}, the only non-zero components of $\mathring{\omega}_{a\mu\nu}$ are $\mathring{\omega}_{0 0 i}\approx G\partial_i h_{00}$ and $\mathring{\omega}_{i j k}\approx G\partial_k h_{ij}$. This implies that each $\Gamma_a$ matrix is proportional to
one of the $4\times 4$ matrices
\begin{equation}
\gamma^{0i}=
\begin{bmatrix}
    0 & \sigma^i\\
    \sigma^i & 0
\end{bmatrix} \quad\text{and}\quad \gamma^{ij}=-\ui{\epsilon^{ij}}_k
\begin{bmatrix}
    \sigma^k & 0\\
    0 & \sigma^k
\end{bmatrix},
    \label{hgamma}
\end{equation}
where $\epsilon_{ijk}$ is the completely anti-symmetric tensor with $\epsilon_{123}=1$.

We further assume that the beam of spin-$\frac{1}{2}$ particles is non-relativistic (cf. Refs.~\cite{Nonrel1,Nonrel2,Nonrel3}). To obtain the corresponding non-relativistic equation, we use the well-known Foldy-Wouthuysen~\cite{fw} procedure, namely, a series of unitary spinor transformations that block-diagonalizes the Hamiltonian $\tilde H$ order by order in $p_i/m$, where $p_i$ are the components of momentum. When we multiply Eq.~\eqref{diraco} on the left by $\gamma^0$, we get $\ui\partial_t\psi=\tilde H\psi$~\cite{Yuri} for
\begin{eqnarray}
\tilde H=\gamma^0 m+\mathcal E+\mathcal O ,
    \label{nrhtilde}
\end{eqnarray}
where
\begin{equation}
    \mathcal E=\frac{3\pi G}{8}J_5^i\gamma^0\gamma_5\gamma_i
    \label{even}
\end{equation}
and
\begin{equation}
    \mathcal O=-\ui\gamma^0\Gamma_0-\ui\gamma^0\gamma^i\partial_i-\ui\gamma^0\gamma^i\Gamma_i+\frac{3\pi G}{8}J_5^0\gamma_5.
    \label{odd}
\end{equation}
Here, $\mathcal E$ and $\mathcal O$ are even and odd terms, respectively, according to the Foldy-Wouthuysen terminology\footnote{An even matrix is block diagonal as a Dirac matrix while an odd matrix is a non-zero matrix with zero block diagonal entries. The product of two even or two odd matrices is even; an odd matrix times an even matrix is odd.}. To the lowest order in $p_i/m$, the block diagonal Hamiltonian $\tilde H^\prime$, obtained from a Foldy-Wouthuysen similarity transformation from $\tilde H$, has the form~\cite{fw}
\begin{eqnarray}
\tilde H'\approx \gamma^0 m+\mathcal E+\frac{\gamma^0}{2m}\mathcal O^2,
\end{eqnarray}
and the particle (anti-particle) $2 \times 2$ Hamiltonian corresponds to the upper (lower) block, which we use to write a Shr\"odinger-like equation for a two-component spinor.

By inspection, we can conclude that all contributions containing $\Gamma_a$ enter $\tilde H$ as a multiple of the identity matrix, which commutes with all spin operators. Therefore, to linear order in $G$, the interactions mediated by the metric cannot mix the spinorial components of the beam. However, this is not the case for torsion. Thus, given that we are primarily interested in distinguishing EC from GR by measuring changes in spin, we omit the metric interactions until we reintroduce them phenomenologically later.

Hence, the relevant non-relativistic Hamiltonian for a particle travelling in the $+x$ direction, to first order in $p_i/m$ and $G$, is
\begin{equation}
    H=\frac{p^2}{2m}+\frac{3\pi G}{8}\left[\frac{\sigma_x J_5^0p}{2m}+\frac{\sigma_x p(J_5^0\cdot)}{2m}-\sum_{j=1}^3 (J_5)^j\sigma_j\right],
    \label{hamil}
\end{equation}
where $p\equiv p_x= -\ui \partial_x$. In Eq.~\eqref{hamil}, the symbol $\cdot$ emphasizes that $p$ also acts on the vector on which $H$ operates. Recall that, in the non-relativistic limit, there is a preferred notion of time, which explains $J_5^0$ and $J_5^i$ playing different roles.

The simplest model for the target capable of distinguishing EC from GR consists of a space-like axial current whose components are constant inside the target and are zero outside. Such a current is incompatible with a free Dirac equation for a massive $\Psi$ field. This is because spinors inside a target interact with its constituents; hence, we do not impose the free Dirac equations on $\Psi$.

In the present form, the Hamiltonian~\eqref{hamil} cannot accurately represent a physical setting because it neglects the $\mathcal O(G)$-gravitational interactions encoded in the metric. As we saw above, these interactions could be modelled using Newtonian gravity and are proportional to the $2\times 2$ identity matrix, $\mathbf 1$. However, since our goal is to obtain an order-of-magnitude estimate for the spin polarization, we simplify the potential to a step function, $V_0^g\theta(x)\theta(a-x)\mathbf 1$ for a constant $V_0^g<0$ and where $\theta$ is the Heaviside step function. Moreover, to account for the known short-ranged spin-independent interactions between the target and the neutron beam, we add a potential to the Hamiltonian~\eqref{hamil} of the same form. The potential is  $V_0\theta(x)\theta(a-x)\mathbf 1$ for a constant $V_0>0$ because the repulsive short-ranged interactions are several orders of magnitude more intense than the gravitational attraction between target and beam. This potential should, in principle, be derived from first principles, but this can only be achieved by considering an interacting Dirac field and integrating all interactions, which lies outside the scope of the present paper. We turn to find some physical consequences of the Hamiltonian~\eqref{hamil}.

\section{Observable implications}

To find physical consequences, we solve the stationary Schr\"odinger equation for the particle Hamiltonian~\eqref{hamil}. We have in mind an experiment involving polarized neutrons. Slowly moving neutrons are natural candidates for the beam particles as they avoid Coulomb-type interactions and have long been used as probes of fundamental physics~\cite{SlowNeutronsReview}. They can also be handled and measured with exquisite sensitivities (see, e.g., Refs.~\cite{Andreev2018, Libertad, PhysRevC.100.015204, PhysRevLett.124.081803, PhysRevLett.125.131803}; Ref.~\cite{Whitepaper} is a review), and experiments where the spin polarization of a neutron beam that goes through a polarized media have already been performed~\cite{Angle3}.

The solutions to the stationary Schr\"odinger equation are found in \ref{App} for a non-relativistic monochromatic beam of neutrons with polarization along $+z$~\cite{Monochromatic} directed towards a spin-polarized target whose geometry is described above (see Fig.~\ref{fig}). In this Appendix, we solve the Schr\"odinger equation in three regions: $x<0$, $0<x<a$, and $x>a$. Then, the solutions are ``glued'' by requiring continuity for the wavefunction $\psi$ and fixing the discontinuity of $\partial_x \psi$ at $x=0$ and $x=a$ by integrating the Schr\"odinger equation, which contains Dirac deltas, around those regions.

We write the spinor at $x<0$ as $\psi=\psi_i+\psi_r$, where the first term contains the incoming right-propagating wave and the second is the reflected part of the wavefunction. Then, we can obtain the reflection coefficient from the ratio between the reflected and incident beams. Analogously, at $x>a$ we denote $\psi=\psi_t$. We define $\phi_r$ and $\phi_t$ by
\begin{equation}
\cos\phi_r=\frac{\psi_r^\dagger\vec\sigma\psi_r\cdot\psi_i^\dagger\vec\sigma\psi_i}{|\psi_r^\dagger\vec\sigma\psi_r|\ |\psi_i^\dagger\vec\sigma\psi_i|},\quad
\cos\phi_t=\frac{\psi_t^\dagger\vec\sigma\psi_t\cdot\psi_i^\dagger\vec\sigma\psi_i}{|\psi_t^\dagger\vec\sigma\psi_t|\ |\psi_i^\dagger\vec\sigma\psi_i|}.
    \label{transangleamp}
\end{equation}
These angles characterize the change of the expectation value of the spin polarization for the reflected and transmitted beams, respectively, as compared with the incident beam. Here, the dagger denotes the transpose conjugated. A method to find the explicit form of the corresponding parts of the wavefunction is described in \ref{App}. Relevantly, even though these last equations are ratios containing terms up to $\mathcal{O}(G^2)$, the result is linear in $G$.

The resulting expressions for $\phi_r$ and $\phi_t$ are lengthy, but we can consider the limit when the thickness $a$ of the target is much larger than the real and imaginary parts of all wavelengths. We get, for the case where $V_0>E$, that
\begin{widetext}
\begin{eqnarray}
    \phi_r&\approx&\frac{3\pi G}{4}\frac{k_5 m |J_5^x+J_5^y+J_5^z|}{\kappa(\kappa^2+k_5^2)},
    \label{reflecinfinity}\\
    \phi_t &\approx& \frac{3\pi Gma}{8\kappa}\sqrt{\vec{J_5}^2+2(J_5^xJ_5^y+J_5^yJ_5^z+J_5^xJ_5^z)+\frac{4(J_5^0)^2\kappa^2}{m^2}}\label{transinfty},  
\end{eqnarray}
\end{widetext}
where $k_5\equiv\sqrt{2mE}$,  $\kappa\equiv\sqrt{2m(V_0-E)}$, and $\vec{J_5}^2\equiv (J_5^x)^2+(J_5^y)^2+(J_5^z)^2$. Importantly, given that GR is recovered by setting $J_5^\mu =0$, it is possible to verify that, according to GR, there is no ``spin deflection'' to order $\mathcal{O}(G)$. Recall that, although there are $\mathcal{O}(G)$ spin-independent gravitational effects inside and outside the target [see Eq.~\eqref{linconn}], they are absorbed in $V_0$ in our estimates, together with all the other spin-independent interactions. We also computed the case when $E> V_0$, even though it is less useful when considering ultra-slow neutrons. The results for the reflected and transmitted angles are similar to Eqs.~\eqref{reflecinfinity}-\eqref{transinfty}, and the estimated effects are of the same order of magnitude. 

To obtain order-of-magnitude estimations for $\phi_r$ and $\phi_t$, we consider a realistic~\cite[Sec. 2.3.1.]{SlowNeutronsReview} speed for the ultra-cold neutrons of $5\mathrm{ms}^{-1}$. Moreover, we consider the target polarized along the $+y$ direction. Importantly, there exist spin-polarized targets that are insensitive to magnetic effects~\cite{Adelberger1, Adelberger12}. These targets have an effective number of polarized particles of $10^{23}$ in a cylindrical device whose diameter and height are roughly $5\mathrm{cm}$, resulting in a density of spin-polarized particles of $2.5\cdot 10^{26}\mathrm{m}^{-3}$, which is the considered density, and we take $J_5^0\to 0$\footnote{When $J_5^0=0$, the torsion effect looks like that of an external magnetic field $B^i=3\pi GJ_5^i/4$. With our numerical estimates and when $J^0_5$ is in the range $[0,J^y_5)$, the result given in Eq.~\eqref{transinfty} is only sensitive to $J^0_5$ in the $15^\mathrm{th}$ decimal digit. One of the lessons of the present study is that, if one wishes to single out the peculiar dependence on $J^0_5$ and $p$ given in the second and third terms of the Hamiltonian \eqref{hamil}, one must use a non-zero $J_5^0$ and seek parameters such that $8(V_0-E)\gg m$.}. We also set $V_0\approx 1.5 E$. Then, Eq.~\eqref{reflecinfinity} yields $\phi_r\approx 5.5\cdot 10^{-47}$ radians, and Eq.~\eqref{transinfty} produces $\phi_t\approx (3.3\cdot 10^{-36})(a/\mathrm{m})$. Note that the transmitted angle is proportional to $a$, implying that this effect is cumulative: the angle grows with the target's thickness. Of course, it can also be enlarged by manipulating the beam to go through the target several times; the physical limitation in this case is the neutron's lifetime. These estimations ought to be compared with the achieved sensitivity for the neutron's polarization angle, which has been measured at the considerable rate of $10^{-7}$ radians per metre~\cite{Angle, Angle2, Angle3}. Thus, according to these estimations, we are still some $20$ orders of magnitude away from any possible torsion detection. However, things improve significantly by fine-tuning the physical parameters.

Naively, one could think that the limit where torsion effects dominate is when $V_0\to 0$. However, as can be seen by inspecting Eqs.~\eqref{reflecinfinity}-\eqref{transinfty}, the relevant limit is when the potential $V_0$ approaches $E$, and hence, $\kappa \to 0$. We take this limit before setting $a$ much longer than all wavelengths. The results are
\begin{eqnarray}
\phi_r&\approx&\frac{\pi G}{2}\frac{ma|J_5^x+J_5^y+J_5^z|}{k_5} \sim 3\cdot10^{-36}\ \left(\frac{a}{\mathrm{m}}\right),
    \label{refk0}\\
\phi_t&\approx&\frac{\pi G}{8}|J_5^x+J_5^y+J_5^z|ma^2\sim 6\cdot 10^{-26}\ \left(\frac{a}{\mathrm{m}}\right)^2,
    \label{transk0}
\end{eqnarray}
where we used the values of the previous paragraphs to get numerical estimates. Surprisingly, the size of the effects is roughly $10$ orders of magnitude larger than in the case where $\kappa \approx 1$, and the angles also acquire an extra factor of $a$. Achieving this limit is expected to be complicated, although, in ideal situations, an experimentalist could tune $E$ so that $\phi_r$ and $\phi_t$ are maximized. Still, we hope that, eventually, in a situation where $\kappa\approx 0$, with better sensitivity in the measurement of the polarization angle, and using targets that are larger and that have higher spin density, we could settle the debate between GR and EC.

\section{Final remarks}

In this Letter, we show that EC predicts a shift in the reflected and transmitted neutron polarization that is linear in $G$, which is absent in GR. Thus, if this shift is measured, it would constitute a smoking gun for torsion. Of course, other interactions could generate similar spin shifts~\cite{Dobrescu}. However, the effects of EC theory could eventually be identified by their dependence on the beam and target parameters. Furthermore, EC introduces no new fundamental constants for which bounds could be set.

The proposal presented here shares several features with the tests of local Lorentz invariance. In the latter, the precision in some parameters was improved by several orders of magnitude in a few decades, reaching Planck scale sensitivity (a comprehensive list of limits on local Lorentz invariance is found in Ref.~\cite{Datatables}). In both cases, the sought effects are cumulative, amplified at given energies, and involve high-precision experimental techniques. A realistic experiment to look for torsion will likely require more detailed modelling of the target. Still, the observables proposed here have the potential to experimentally distinguish GR and EC, which would have far-reaching implications.

\appendix \label{Appendix}
\section{Reflection and transmission problem}\label{App}

In this appendix, we solve the stationary Schr\"odinger equation, $H\psi=E\psi$, for the particle Hamiltonian \eqref{hamil}, considering a monochromatic beam of neutrons with momentum along the $x$ direction, and whose incident polarization is $+z$, and for the target described above. For this purpose, we use standard methods~\cite{teschl12} to convert the second-order Schr\"odinger equations as a first-order linear system of equations for $\psi=\binom{f}{g}$ and its associated momenta $-\ui \partial_xf\equiv p_f$ and $-\ui \partial_xg \equiv p_g$. We use the spinor basis where $\sigma_3$ is diagonal and, for compactness, matrix notation. The system of equations takes the form
\begin{equation}
\frac{\ud}{\ud x}
\begin{bmatrix}
    f\\
    g\\
    p_f\\
    p_g
\end{bmatrix}
=A(x)
\begin{bmatrix}
    f\\
    g\\
    p_f\\
    p_g
\end{bmatrix},
\label{diracnr}
\end{equation}
where $A(x)$ is defined as
\begin{equation*}
A=    \begin{bmatrix}
    \mathbf{0}_2 & \ui \mathbf{1}_2 \\
    2m \mathbf B -J_5^0 \mathbf C & \mathbf D
\end{bmatrix},
\end{equation*}
with
\begin{equation*}
\mathbf B=    \begin{bmatrix}
   \ui\left(E-V_0+\frac{3\pi G}{8}J_5^z\right) & \frac{3\pi G}{8} \left(\ui J_5^x+J_5^y\right) \\
   \frac{3\pi G}{8} \left(\ui J_5^x-J_5^y\right)& \ui\left(E-V_0+\frac{3\pi G}{8}J_5^z\right) 
\end{bmatrix},
\end{equation*}
\begin{equation*}
\mathbf C=   \frac{3\pi G}{8} \begin{bmatrix}
   0 &\delta(x)-\delta(x-a) \\
    \delta(x)-\delta(x-a) & 0
\end{bmatrix},
\end{equation*}
and
\begin{equation*}
\mathbf D=   -\ui \frac{3\pi G}{4}J_5^0   \begin{bmatrix}
     0 & 1\\
     1 & 0
\end{bmatrix}.
\end{equation*}
The Dirac deltas in $\mathbf C$ arise from Eq.~\eqref{hamil} once $p$ acts on $J_5^0$.

We find solutions in each spatial sector and then impose boundary conditions. At $x<0$, we take the solution as a right-moving spinor polarized in the $+z$ direction, which represents the incident spinor, plus a reflected spinor with arbitrary polarization, yielding
\begin{equation}
  \begin{aligned} \label{leftansatz}
    f(x) &= \ue^{\ui k_5 x}+b\ue^{-\ui k_5x}, \quad x<0, \\
    g(x) &= c\ue^{-\ui k_5x}, \quad\quad\quad\quad x<0,
  \end{aligned}
\end{equation}
where $k_5\equiv\sqrt{2mE}$ and $b$ and $c$ are complex constants to be determined. The overall normalization is immaterial, and we exploit it to fix the coefficient in the incident part of the wavefunction (the coefficient in the first term of $f$). Similarly, in $x>a$, we only consider right-moving waves:
\begin{equation}
  \begin{aligned} \label{rightansatz}
    f(x) &=  h\ue^{\ui k_5x},\quad\quad\quad\quad  x>a, \\
    g(x) &=r\ue^{\ui k_5x},\quad\quad\quad\quad  x>a,
  \end{aligned}
\end{equation}
where $h$ and $r$ are complex amplitudes that are fixed below. One can readily verify that Eqs.~\eqref{leftansatz} and~\eqref{rightansatz} solve Eq.~\eqref{diracnr}.

Inside the target ($0< x< a$), the matrix $A$ admits four eigenvalues, which we denote by $\ui k_n$, $n=1,2,3,4$. Their corresponding eigenvectors are designated by $\binom{\psi}{-\ui\partial_x\psi}$ with
\begin{equation}
    \psi=\sum_{n=1}^4 d_n Y_n \ue^{\ui k_n x},
\end{equation}
for certain amplitudes $d_n$. The vectors $Y_n$ are found to be, up to normalization,
\begin{equation}
Y_1=Y_2=
\begin{bmatrix}
    -1\\
    1
\end{bmatrix}
\quad\text{and}\quad Y_3=Y_4=
\begin{bmatrix}
    1\\
    1
\end{bmatrix}.
    \label{yjs}
\end{equation}

The eigenvalues are more conveniently expressed in terms of $k_5$ and $\kappa\equiv\sqrt{2m(V_0-E)}$. In what follows, we assume $\kappa$ to be real and positive ($V_0>E$); the case where $\kappa$ is purely imaginary is solved similarly. 
To the lowest order in $G$, these eigenvalues are
\begin{subequations}\label{wavenumbers}
\begin{eqnarray}
k_1&\approx&-\ui\kappa+\frac{3\pi G}{8}\left(J_5^0-\ui\frac{m (J_5^x+J_5^y+J_5^z)}{2\kappa}\right) ,\\
k_2&\approx&\ui\kappa+\frac{3\pi G}{8}\left(J_5^0+\ui\frac{m (J_5^x+J_5^y+J_5^z)}{2\kappa}\right) ,\\
k_3&\approx&-\ui\kappa -\frac{3\pi G}{8}\left(J_5^0 -\ui\frac{m (J_5^x+J_5^y+J_5^z)}{2\kappa}\right), \\
k_4&\approx&\ui\kappa-\frac{3\pi G}{8}\left(J_5^0+\ui\frac{m(J_5^x+J_5^y+J_5^z)}{2\kappa}\right).
\end{eqnarray}
\end{subequations}

To fix $b,c,\{d_n\}_{n=1,2,3,4},h,r$, we impose the continuity of $\psi$ across $x=0$ and $x=a$, leading to four algebraic equations. We attain two others by integrating Eq.~\eqref{diracnr} from $-\epsilon$ to $+\epsilon$ and taking the limit $\epsilon\to 0^+$. The remaining two are similarly obtained by integrating Eq.\eqref{diracnr} from $a-\epsilon$ to $a+\epsilon$ and taking the same limit. The resulting equations form a linear system of equations that can be compactly written as
\begin{widetext}
\begin{equation}
    \begin{bmatrix}
        0 & 0 & k_1\ue^{\ui k_1 a} & k_2\ue^{\ui k_2 a} & -k_3\ue^{\ui k_3 a} & -k_4\ue^{\ui k_4 a} & k_5\ue^{\ui k_5 a} & -\frac{3\pi}{8} GJ_5^0\ue^{\ui k_5a}\\
        0 & 0 & -k_1\ue^{\ui k_1 a} & -k_2\ue^{\ui k_2a} & -k_3\ue^{\ui k_3a} & -k_4\ue^{\ui k_4a} & -\frac{3\pi}{8} GJ_5^0\ue^{\ui k_5a} & k_5\ue^{\ui k_5 a}\\
        \frac{3\pi}{8} GJ_5^0 & k_5 & k_1 & k_2 & k_3 & k_4 & 0 & 0\\
        k_5 & \frac{3\pi}{8} GJ_5^0 & -k_1 & -k_2 & k_3 & k_4 & 0 & 0\\
        0 & 0 & \ue^{\ui k_1a} & \ue^{\ui k_2a} & \ue^{\ui k_3a} & \ue^{\ui k_4a} & 0 & -\ue^{\ui k_5a}\\
        0 & 0 & -\ue^{\ui k_1a} & -\ue^{\ui k_2a} & \ue^{\ui k_3a} & \ue^{\ui k_4a} & -\ue^{\ui k_5a} & 0\\
        -1 & 0 & -1 & -1 & 1 & 1 & 0 & 0\\
        0 & -1 & 1 & 1 & 1 & 1 & 0 & 0
    \end{bmatrix}.
    \begin{bmatrix}
    b\\
    c\\
    d_1\\
    d_2\\
    d_3\\
    d_4\\
    h\\
    r
\end{bmatrix}
=
\begin{bmatrix}
    0\\
    0\\
    -\frac{3\pi}{8} GJ_5^0\\
    k_5\\
    0\\
    0\\
    1\\
    0
\end{bmatrix}.
\label{linsys}
\end{equation}
\end{widetext}
Note that because the incoming beam has only spin-up components, any non-zero $c$ or $r$, which represent spin-down components in the reflected and transmitted beams, respectively, attest to a spin-dependent effect, which, in our case, is due to torsion.

\section*{Acknowledgments}
We acknowledge exchanges with W. G. Snow, C. Chryssomalakos, and D. Sudardsky and the financial support from CONACyT through the grant FORDECYT-PRONACES 140630. BAC receives funding from a UNAM-DGAPA fellowship.

\bibliography{references}

\end{document}